\documentclass{appolb}
\usepackage{epsfig}

\begin{document}
\pagestyle{plain}
\newcount\eLiNe\eLiNe=\inputlineno\advance\eLiNe by -1
\title{Neutrinos as Hot or Warm Dark Matter 
\thanks{Presented by Zhi-zhong Xing at the XXXV International Conference of Theoretical
Physics --- ¡°Matter to the Deepest¡±, Ustron, Poland, September
12--18, 2011.}%
}
\author{Y.F. Li and Zhi-zhong Xing
\address{Institute of High Energy Physics and Theoretical Physics
Center for Science Facilities, Chinese Academy of Sciences, Beijing
100049, China}} \maketitle

\begin{abstract}
Both active and sterile sub-eV neutrinos can serve for hot dark
matter (DM). On the other hand, keV sterile neutrinos could be a
good candidate for warm DM. The beta-decaying (e.g.,$^3{\rm H}$ and
$^{106}{\rm Ru}$) and EC-decaying (e.g., $^{163}{\rm Ho}$) nuclei
are considered as the most promising targets to capture those
extremely low energy neutrinos and antineutrinos, respectively. We
calculate the capture rates of relic electron neutrinos and
antineutrinos against the corresponding beta-decay or EC-decay
backgrounds in different flavor mixing schemes. We stress that such
direct laboratory measurements of hot or warm DM might not be
hopeless in the long term.
\end{abstract}

\section{Introduction}

Although the existence of dark matter (DM) in the Universe has been
established, what it is made of remains a fundamental puzzle
\cite{XZ}. Within the standard model (SM) three kinds of active
neutrinos and their antiparticles, whose masses lie in the sub-eV
range, may constitute hot DM. Beyond the SM one or more species of
sterile neutrinos and antineutrinos at a similar mass scale may also
form hot DM, if they were thermalized in the early Universe as their
active counterparts. Such light sterile particles are hypothetical,
but their existence is more or less implied by current experimental
and cosmological data. On the one hand, the long-standing LSND
antineutrino anomaly \cite{LSND}, the more recent MiniBooNE
antineutrino anomaly \cite{M} and the latest reactor antineutrino
anomaly \cite{R} can all be interpreted as the active-sterile
antineutrino oscillations in the assumption of two kinds of sterile
antineutrinos whose masses are close to 1 eV \cite{Schwetz}. On the
other hand, an analysis of the existing data on the cosmic microwave
background (CMB), galaxy clustering and Type-Ia supernovae favors
some extra radiation content in the Universe and one or two species
of sterile neutrinos and antineutrinos at the sub-eV mass scale
\cite{Raffelt}. We are therefore open-minded to conjecture that hot
DM might in general consist of both active and sterile components.
These relics of the Big Bang form the unseen cosmic neutrino
background (C$\nu$B) and cosmic antineutrino background
(C$\overline{\nu}$B), whose temperature $T^{}_\nu$ is slightly lower
than the CMB temperature $T^{}_\gamma$ (i.e., $T^{}_\nu =
\sqrt[3]{4/11} \ T^{}_\gamma \simeq 1.945$ eV today) \cite{PDG}.

But hot DM only has a tiny contribution to the total matter density
of the Universe. A careful analysis of the structure formation
indicates that most DM should be cold (nonrelativistic) or warm
(semirelativistic) at the onset of the galaxy formation, when the
temperature of the Universe was about 1 keV \cite{PDG}. A number of
candidates for cold DM have so far been investigated. In comparison,
warm DM is another interesting possibility of accounting for the
observed non-luminous and non-baryonic matter content in the
Universe. It may allow us to solve or soften several problems that
we have recently encountered in the DM simulations \cite{Bode}
(e.g., to damp the inhomogeneities on small scales by reducing the
number of dwarf galaxies or to smooth the cusps in the DM halos). A
good candidate for warm DM should be sterile neutrinos and
antineutrinos, if their masses are in the keV range and their
lifetimes are much longer than the age of the Universe
\cite{Review}.

How to detect hot or warm DM neutrinos and antineutrinos is a great
challenge to the present experimental techniques. Among several
possible ways \cite{Ringwald}, the most promising one is the relic
neutrino capture experiment by means of radioactive beta-decaying
nuclei \cite{Weinberg}--\cite{Vega}. The key point is that a generic
neutrino capture reaction will take place with no threshold on the
incident neutrino energy, provided the mother nuclei can naturally
undergo the beta decay with an energy release in the limit of
vanishing neutrino masses. The signal of this neutrino capture
process is measured by the monoenergetic electron's kinetic energy
for each neutrino mass eigenstate, well beyond that of the
corresponding beta decay background. A measurement of the gap
between the capture and decay processes will directly probe these
hot or warm DM neutrinos and determine or constrain their masses and
mixing angles. However, this method does not directly apply to the
capture of hot or warm DM antineutrinos, simply because it is
$\nu^{}_e$ (instead of $\overline{\nu}^{}_e$) that is involved in
the capture reaction. A possible way out for relic antineutrino
detection is to make use of some radioactive nuclei which can decay
via electron capture (EC) \cite{Cocco2}--\cite{LXJCAP}.

The remaining parts of this talk are organized as follows. In
section 2 we summarize the main formulas which can be used to
calculate the beta-decay energy spectrum and the relic neutrino
capture rate. Section 3 is devoted to the capture of hot or warm DM
neutrinos in different flavor mixing schemes. We shall give a brief
description of the relic antineutrino capture on EC-decaying nuclei
in section 4, and then conclude in section 5.

\section{Relic Neutrino Captures}

In the presence of $3+N^{}_s$ species of active and sterile
neutrinos, the flavor eigenstates of three active neutrinos can be
written as \cite{XZ,PDG}
\begin{equation}
\left|{\nu^{}_\alpha}\right\rangle = \sum_{i} V^*_{\alpha i} \left|
{\nu^{}_i} \right\rangle \; ,
\end{equation}
where $\alpha$ runs over $e$, $\mu$ and $\tau$, $\nu^{}_i$ is a mass
eigenstate of active (for $1 \leq i \leq 3$) or sterile (for $4 \leq
i \leq 3 + N^{}_s$) neutrinos, and $V^{}_{\alpha i}$ stands for an
element of the $3 \times (3+ N^{}_s)$ neutrino mixing matrix
$V^{}_{}$. For simplicity, we assume that the light sterile
neutrinos under consideration do not significantly affect the values
of two mass-squared differences and three mixing angles of active
neutrinos extracted from current experimental data on solar,
atmospheric, reactor and accelerator neutrino oscillations
\cite{PDG}. In this assumption we shall use $\Delta m^2_{21} \approx
7.6 \times 10^{-5} ~{\rm eV}^2$ and $|\Delta m^2_{31}| \approx 2.4
\times 10^{-3} ~{\rm eV}^2$ together with $\theta^{}_{12} \approx
34^\circ$ and $\theta^{}_{13} \approx 10^\circ$ as typical inputs in
our numerical estimates. Depending on the sign of $\Delta m^2_{31}$,
there are two possible mass patterns for active neutrinos: $m^{}_1 <
m^{}_2 < m^{}_3$ (normal hierarchy) or $m^{}_3 < m^{}_1 < m^{}_2$
(inverted hierarchy). In either case the absolute mass scale is
unknown, but its upper bound is expected to be of ${\cal O}(0.1)$ eV
as constrained by current cosmological data \cite{WMAP10}. We shall
specify the values of $m^{}_i$ and $|V^{}_{ei}|$ when calculating
the capture rates of relic neutrinos in sections 3.

Let us concentrate on the relic neutrino capture on radioactive
beta-decaying nuclei (i.e., $\nu^{}_e + N \to N^\prime + e^-$), This
capture reaction can happen for any kinetic energy of the incident
neutrino, because the corresponding beta decay $N \to N^\prime + e^-
+ \overline{\nu}^{}_e$ always releases some energies ($Q^{}_\beta =
m^{}_N - m^{}_{N^\prime} - m^{}_e > 0$). So it has a unique
advantage in detecting cosmic neutrinos with both $m^{}_i \ll
Q^{}_\beta$ and extremely low energies \cite{Weinberg}--\cite{Vega}.
In the low-energy limit the product of the cross section of
non-relativistic neutrinos $\sigma^{}_{\nu^{}_i}$ and the neutrino
velocity $v^{}_{\nu^{}_i}$ converges to a constant value
\cite{Cocco}, and thus the differential neutrino capture rate reads
\begin{equation}
\frac{{\rm d} \lambda^{}_{\nu}}{{\rm d} T^{}_{e}} = \sum_i
|V^{}_{ei}|^2 \sigma^{}_{\nu^{}_i} v^{}_{\nu^{}_i}
n^{}_{\nu^{}_i}\,R(T^{}_{e}, \, T^{i}_{e}) \; ,
\end{equation}
where $n^{}_{\nu^{}_i}$ denotes the number density of relic
$\nu^{}_i$ neutrinos around the Earth. The standard Big Bang model
predicts $\langle n^{}_{\nu^{}_i} \rangle \approx \langle
n^{}_{\overline{\nu}^{}_i} \rangle \approx 56 ~{\rm cm}^{-3}$ today
for each species of active neutrinos and antineutrinos, and this
prediction is also expected to hold for each species of light
sterile neutrinos and antineutrinos if they could be completely
thermalized in the early Universe. The number density of hot
neutrino and antineutrino DM around the Earth may be more or less
enhanced by the gravitational clustering effect when $m^{}_i$ is
larger than 0.1 eV \cite{Wong}. As for the keV sterile neutrinos, we
assume their number density could account for the total amount of DM
in our Galactic neighborhood. In Eq. (2), $R(T^{}_{e}, \,
T^{i}_{e})$ is a Gaussian energy resolution function defined by
\cite{LLX}
\begin{equation}
R(T^{}_{e}, \, T^{i}_{e}) = \frac{1}{\sqrt{2\pi} \,\sigma}
\exp\left[-\frac{(T^{}_{e} - T^{i}_{e})^2}{2\sigma^2} \right] \; ,
\end{equation}
in which $T^{}_{e}$ is the overall kinetic energy of the electrons
detected in the experiment and $T^{i}_{e} = Q^{}_\beta +
E^{}_{\nu^{}_i}$ is the kinetic energy of the outgoing electron for
each incoming mass eigenstate $\nu^{}_i$. Using $\Delta$ to denote
the experimental energy resolution, we have
$\Delta = 2\sqrt{2\ln 2} \,\sigma \approx 2.35482 \,\sigma$.

The main background of a neutrino capture process is its
corresponding beta decay. The finite energy resolution may push the
outgoing electron's ideal endpoint $Q^{}_\beta - {\rm min}(m^{}_i)$
towards a higher energy region, and hence it is possible to mimic
the desired signal of the neutrino capture reaction. Given the same
energy resolution as that in Eq. (2), we can describe the energy
spectrum of a beta decay as \cite{Weinheimer}
\begin{eqnarray}
\frac{{\rm d} {\lambda}^{}_\beta}{{\rm d}T^{}_e} && =
\int_0^{Q^{}_{\beta}- {\rm min}(m^{}_i)} {\rm d} T^\prime_e \,
\left\{\frac{G^2_{\rm F} \, \cos^2\theta^{}_{\rm C}}{2\pi^3} \,
F\left(Z, E^{}_{e}\right) \, |{\cal M}|^2 E^{}_{e}\sqrt{E^2_e -
m^2_e}   \,  \right .
\nonumber \\
& & \left . \times\left(Q^{}_{\beta} - T^\prime_e\right)\sum^4_{i=1}
\left[ |V^{}_{ei}|^2\sqrt{\left(Q^{}_{\beta}- T^\prime_e \right)^2 -
m_i^2} ~ \Theta\left(Q^{}_{\beta} - T^\prime_e - m^{}_i\right)
\right]\right\}
\nonumber \\
& & \times R\left(T^{}_e, T^\prime_e\right) \; ,
\end{eqnarray}
where $T^\prime_e = E^{}_e - m^{}_e$ is the intrinsic kinetic energy
of the outgoing electron, $F(Z, E^{}_{e})$ denotes the Fermi
function, $|{\cal M}|^2$ stands for the dimensionless contribution
of relevant nuclear matrix elements \cite{Weinheimer}, and
$\theta^{}_{\rm C} \simeq 13^\circ$ is the Cabibbo angle. In Eq. (4)
the theta function $\Theta(Q^{}_{\beta} - T^\prime_e - m^{}_i)$ is
adopted to ensure the kinematic requirement. Note that the numerical
results of $\lambda^{}_{\nu}$ and $\lambda^{}_{\beta}$ can be
properly normalized by using the half-life of the mother nucleus via
the relation ($\lambda^{}_{\beta}T^{}_{1/2}=\ln 2\,$). Then the
distributions of the numbers of signal and
background events are expressed, respectively, as
\begin{eqnarray}
\frac{{\rm d} N^{}_{\rm S}}{{\rm d} T} \hspace{-0.2cm} & = &
\hspace{-0.2cm} {1 \over \lambda^{}_{\beta}} \cdot {{\rm d}
\lambda^{}_{\nu} \over {\rm d} T} \cdot {\ln 2 \over T^{}_{1/2}} \,
N^{}_{\rm T} \, t \; ,
\nonumber \\
\frac{{\rm d} N^{}_{\rm B}}{{\rm d} T} \hspace{-0.2cm} & = &
\hspace{-0.2cm} {1 \over \lambda^{}_{\beta}} \cdot {{\rm d}
\lambda^{}_{\beta}\over {\rm d} T} \cdot {\ln 2 \over T^{}_{1/2}} \,
N^{}_{\rm T} \, t \;
\end{eqnarray}
for a given target factor $N^{}_{\rm T}$ (i.e., the number of target
atoms) and for a given exposure time $t^{}_{}$ in the experiment.

\section{Hot or Warm DM Neutrinos}

We illustrate the relic neutrino capture signals against the
beta-decay backgrounds by considering three neutrino mixing schemes:
(a) the standard scheme with three sub-eV active
neutrinos (hot DM); (b) the (3 + 2) scheme with three sub-eV
active neutrinos and two sub-eV sterile neutrinos (hot DM); and (c) the
(3 + 1) scheme with three sub-eV active neutrinos and one keV
sterile neutrino (warm DM). In our numerical calculations
we typically take 100 g $^3{\rm H}$ as the radioactive target
for the hot DM detection and 10 kg $^3{\rm H}$ or 1 ton
$^{106}{\rm Ru}$ as the isotope source for the warm DM detection.
\begin{figure}[t]
\begin{center}
\begin{tabular}{cc}
\includegraphics*[bb=18 18 280 216, width=0.46\textwidth]{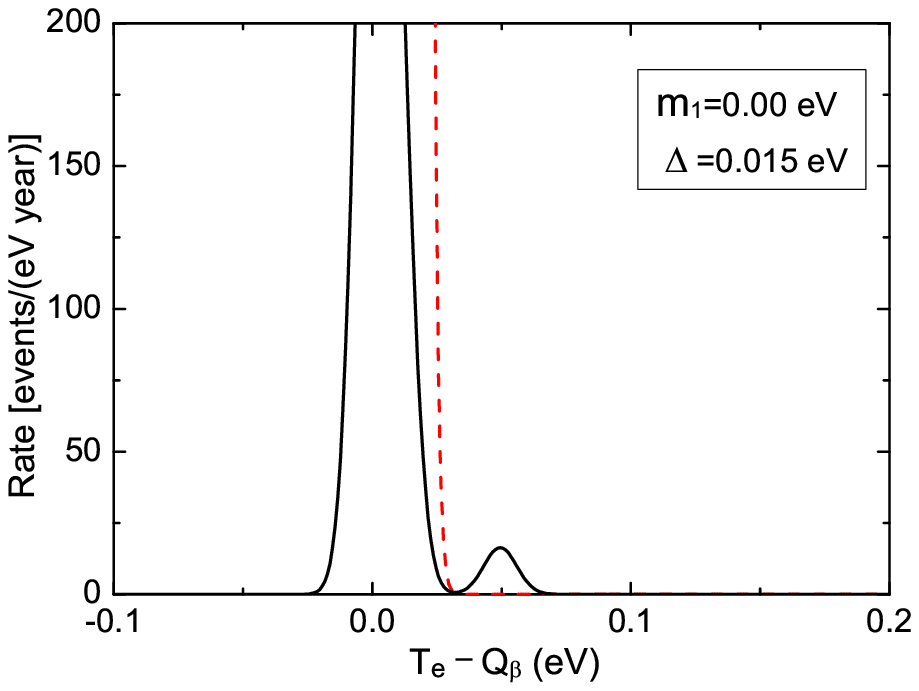}
&
\includegraphics*[bb=18 18 280 216, width=0.46\textwidth]{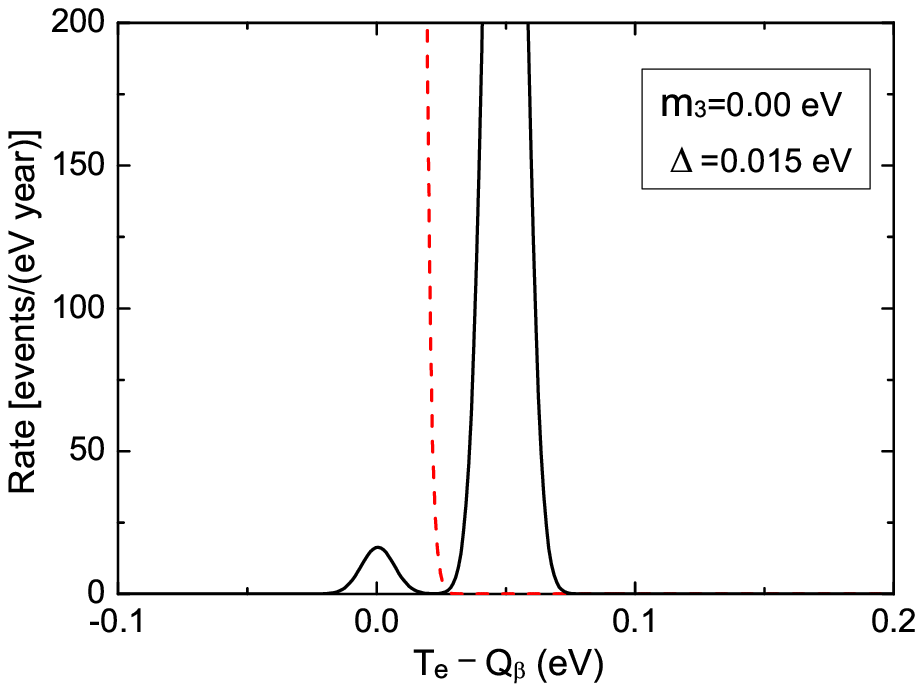}
\\
\includegraphics*[bb=18 18 280 216, width=0.46\textwidth]{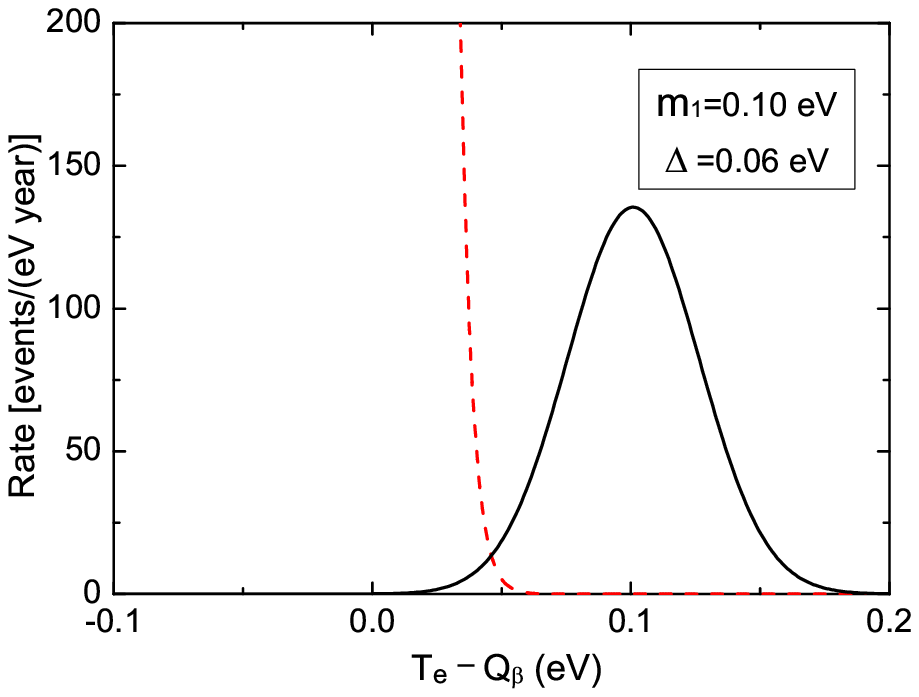}
&
\includegraphics*[bb=18 18 280 216, width=0.46\textwidth]{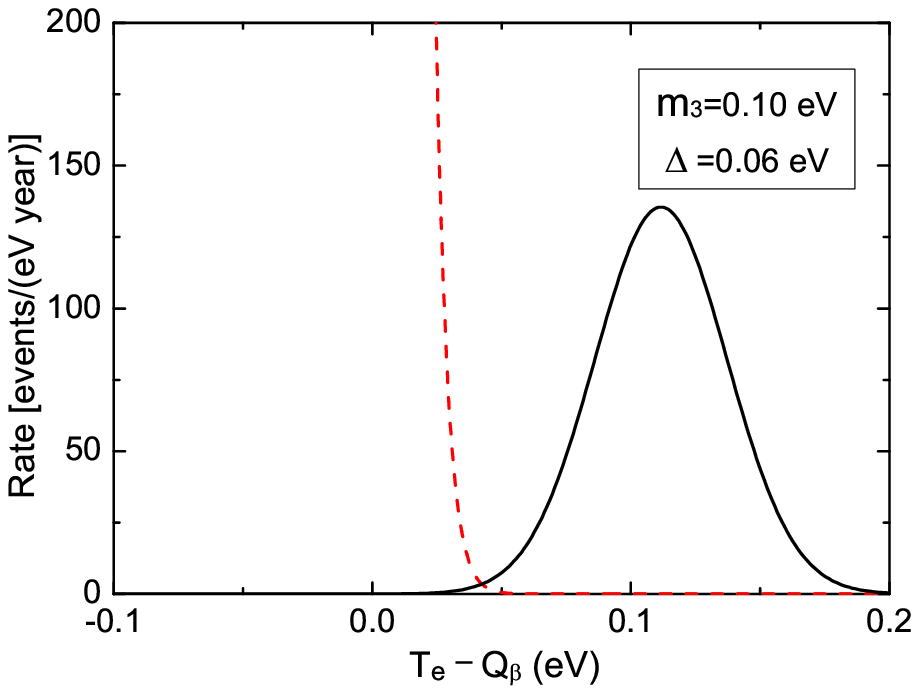}
\end{tabular}
\end{center}
\vspace{-0.5cm}
\caption{The relic neutrino capture rate as a function of the
kinetic energy of electrons in the standard scheme with $\Delta
m^2_{31} > 0$ (left panel) or $\Delta m^2_{31} < 0$ (right panel).}
\end{figure}

Fig. 1 shows the relic neutrino capture rate as a function of the
kinetic energy $T^{}_e$ of electrons in the standard scheme with
$\Delta m^2_{31} > 0$ in the left panel and $\Delta m^2_{31} < 0$ in
the right panel. The finite energy resolution $\Delta$
is taken in such a way that only one single peak can be observed
beyond the background. To illustrate the flavor effects, we fix two
mass-squared differences and three mixing angles with the typical
values given in section 2.
Eq. (2) tells us that the contribution of each
neutrino mass eigenstate $\nu^{}_i$ (for $i=1,2,3$) to the capture
rate is located at $T^{i}_{e} = Q^{}_\beta + E^{}_{\nu^{}_i}$ and
weighted by the relic neutrino number density (we assume
$n^{}_{\nu^{}_i}=\langle n^{}_{\nu^{}_i} \rangle\simeq 56 ~{\rm
cm}^{-3}$) and the flavor mixing matrix element $|V^{}_{ei}|^{2}$.
On the other hand, Eq. (4) tells us that the energy spectrum of the
beta decay near its endpoint is dominated by the lightest neutrino
mass eigenstate hidden in $\nu^{}_{e}$ and sensitive to the energy
resolution $\Delta$. As the smallest neutrino mass ($m^{}_1$ in the
left panel of Fig. 1 or $m^{}_3$ in the right panel of Fig. 1)
increases from 0 to 0.1 eV, the capture signal moves towards the
larger $T^{}_{e}-Q^{}_{\beta}$ region. In comparison, the shift of
the corresponding background is less obvious because the smallest
neutrino mass and $\Delta$ have the opposite effects on the location
of the spectral endpoint of the beta decay. Hence the distance
between the peak of the signal and the background becomes larger for
a larger value of the smallest neutrino mass, and accordingly the
required energy resolution $\Delta$ becomes less stringent.
Comparing between the left and right panels, one can also see that it
is more or less easier to detect relic neutrinos in the $\Delta
m^2_{31} <0$ case, where the signal is separated more obviously from
the background. The reason is simply that the dominant neutrino mass
eigenstates $\nu^{}_1$ and $\nu^{}_2$ have slightly larger
eigenvalues in this case than in the $\Delta m^2_{31} > 0$ case.
\begin{figure}[t]
\begin{center}
\begin{tabular}{cc}
\includegraphics*[bb=18 18 274 212, width=0.46\textwidth]{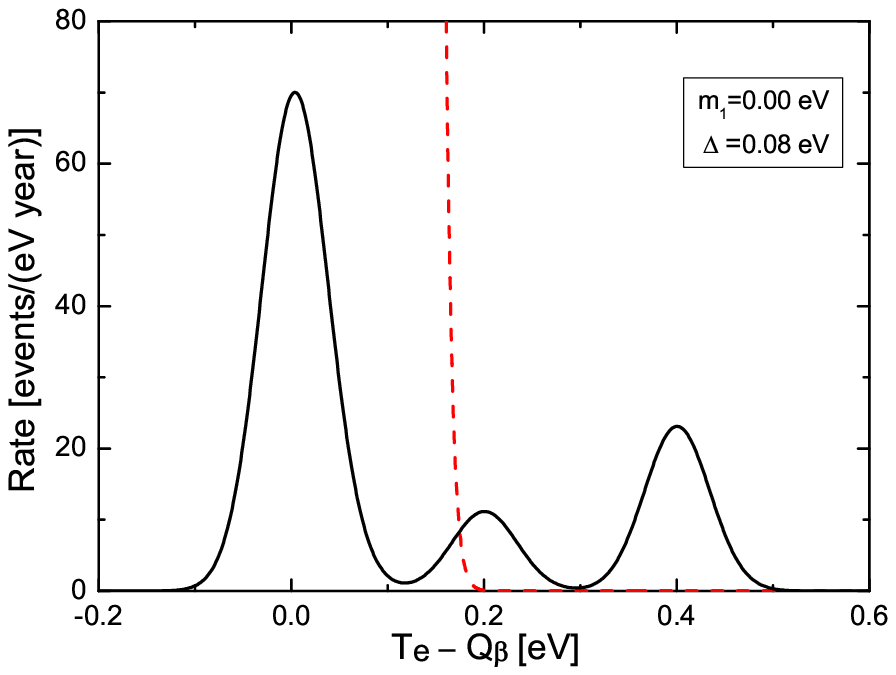}
&
\includegraphics*[bb=18 18 274 212, width=0.46\textwidth]{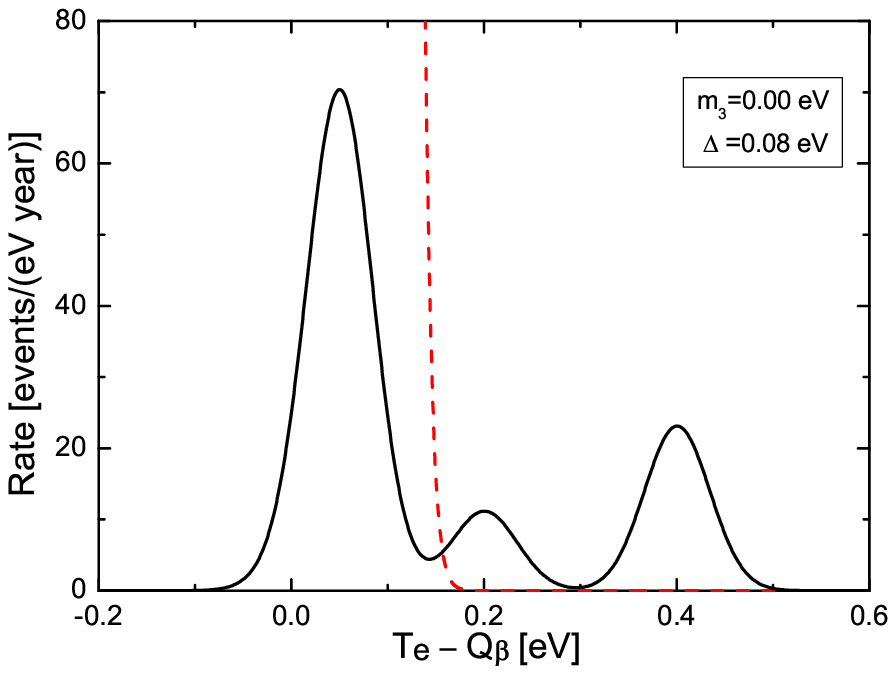}
\end{tabular}
\end{center}
\vspace{-0.5cm}
\caption{The relic neutrino capture rate as a function of the kinetic
energy of electrons in the (3 + 2) scheme with $\Delta m^2_{31} > 0$
(left panel) and $\Delta m^2_{31} < 0$ (right panel). The
gravitational clustering of relic sterile neutrinos around the Earth
has been illustrated by taking $\zeta^{}_1 = \zeta^{}_2 = \zeta^{}_3
=1$ and $\zeta^{}_5 = 2 \zeta^{}_4 = 10$ for example.}
\end{figure}

Now we look at the (3 + 2) scheme with two sub-eV sterile
neutrinos. Considering the preliminary hints of sub-eV sterile
neutrinos \cite{Schwetz,Raffelt}, we simply
assume $m^{}_4 = 0.2$ eV and $m^{}_5 = 0.4$ eV together with
$|V^{}_{e1}| \approx 0.792$, $|V^{}_{e2}| \approx 0.534$,
$|V^{}_{e3}| \approx 0.168$, $|V^{}_{e4}| \approx 0.171$ and
$|V^{}_{e5}| \approx 0.174$ in our numerical estimates. With the
help of Eqs. (2) and (3), we are able to calculate the relic
neutrino capture rate as a function of the kinetic energy $T^{}_e$
of electrons against the corresponding background for both $\Delta
m^2_{31} > 0$ and $\Delta m^2_{31} < 0$ cases in Fig 2. We have taken
$m^{}_1 =0$ or $m^{}_3 =0$ for simplicity. To
illustrate possible gravitational clustering effects, we typically
take $\zeta^{}_1 = \zeta^{}_2 = \zeta^{}_3 =1$ (without clustering
effects for three active neutrinos because their maximal mass is
about 0.05 eV in the scenario under discussion) and $\zeta^{}_5 = 2
\zeta^{}_4 = 10$ (with mild clustering effects for two sterile
neutrinos because their masses are 0.2 eV and 0.4 eV, respectively).
As shown in Fig. 2, the signals of two sterile neutrinos are
obviously enhanced due to $\zeta^{}_4 >1$ and $\zeta^{}_5 >1$. If
the gravitational clustering of non-relativistic neutrinos is very
significant around the Earth, it will be very helpful for us to
detect the relic neutrinos by means of the neutrino capture
processes.

Next let us discuss the detection prospects of keV warm DM
neutrinos. In the (3 + 1) scheme the mass and
mixing element of the keV sterile neutrino are strictly constrained by
current observational data \cite{Review}. Here
we take $m^{}_4 = 2 ~{\rm keV}$ and $|V^{}_{e4}|^2 \simeq 5 \times
10^{-7}$ for illustration, and assume $\Delta m^2_{31} > 0$
and $m^{}_{1}=0$. Our results are given in Fig. 3, where
two different isotope sources (i.e., 10 kg $^3{\rm H}$ and 1 ton
$^{106}{\rm Ru}$) are considered to make a comparison.
We see that it is easy to achieve the required energy resolution.
The main problem which makes the observability
rather dim and remote is the extremely small active-sterile neutrino
mixing angle. We display the half-life effects for both
isotopes in Fig. 3. The finite lifetime is
important for the source of $^{106}{\rm Ru}$ nuclei but negligible
for that of $^3{\rm H}$ nuclei. It may reduce about $30\%$ of
the neutrino capture rate on $^{106}{\rm Ru}$ in the vicinity of
$T^{}_e - Q^{}_\beta \simeq m^{}_4$. Hence this effect must be taken
into account if the duration of such an experiment is comparable
with the half-life of the source.
\begin{figure}[t]
\begin{center}
\begin{tabular}{cc}
\includegraphics*[bb=18 18 275 216, width=0.46\textwidth]{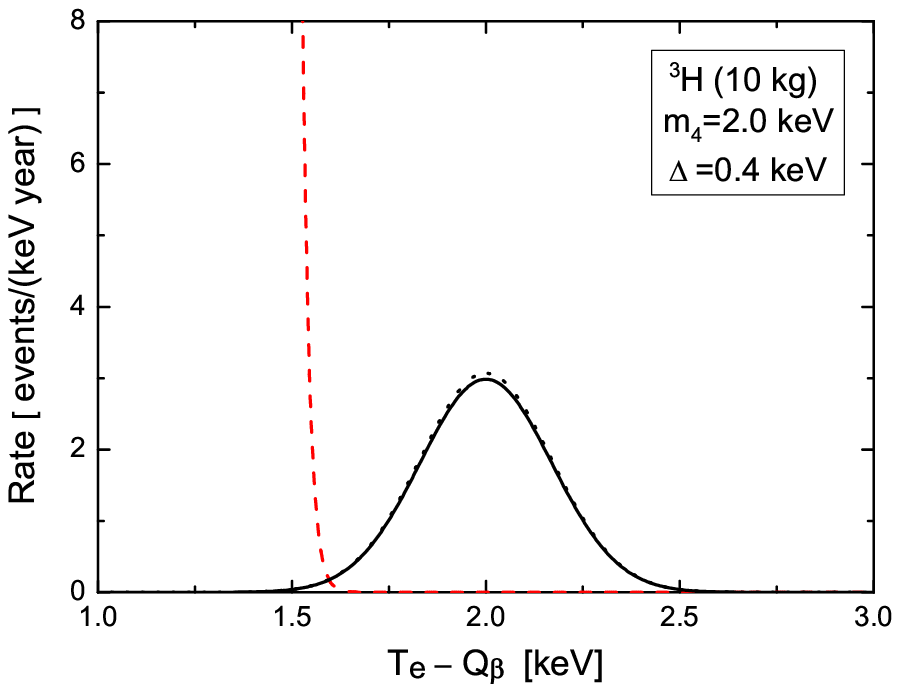}
&
\includegraphics*[bb=18 18 275 216, width=0.46\textwidth]{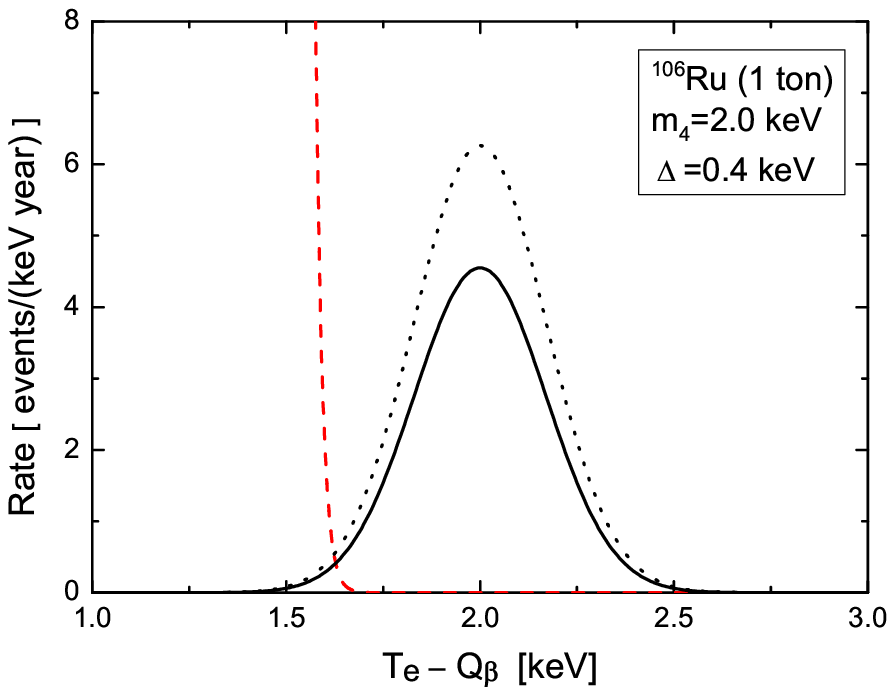}
\end{tabular}
\end{center}
\vspace{-0.5cm}
\caption{The keV sterile neutrino capture rate as a function of the
kinetic energy of electrons with $^3{\rm H}$ (left panel) and
$^{106}{\rm Ru}$ (right panel) as our target sources. The solid (or
dotted) curves denote the signals with (or without) the half-life
effect.}
\end{figure}

\section{Hot or Warm DM Antineutrinos}

As mentioned in section 1, the beta-decaying nuclei can only be used
as the targets of relic neutrino captures and one should employ the
EC-decaying nuclei to detect relic antineutrinos. In this
section we just give a brief discussion about the feasibility of this
method by considering the rather stable isotope $^{163}{\rm Ho}$
\cite{Lusignoli} as our target. We want to emphasis that the fine
structure near the spectral endpoint of the $^{163}{\rm Ho}$ EC
decay could be used to study the lepton flavor effects in a similar
manner as that of beta decays. Comparing between
the left and right panels of Fig. 4 might allow one to
distinguish between the normal and inverted neutrino mass
hierarchies. The properties of the antineutrino capture rate against
its background are more or less the same as those discussed in
section 3. The detection method can in principle be applied to
hot DM (sub-eV active and sterile antineutrinos) and even warm
DM (keV sterile antineutrinos). We
show two numerical examples in Fig. 5 for illustration. The
standard scheme with $\Delta m^2_{31} > 0$ and $m^{}_{1}=0.1~{\rm
eV}$ is depicted in the left panel and a total rate of one event per
year needs 30 kg $^{163}{\rm Ho}$. In the right panel we assume
the existence of a keV sterile antineutrino with the same mass
and mixing angle as the keV sterile neutrino discussed in section 3.
It turns out that we need
as much as 600 ton $^{163}{\rm Ho}$ to get one event per year. So it
is almost hopeless in this scenario. Much more discussions can be
found in our recent works \cite{LX11EC,LXJCAP}.
\begin{figure}[t]
\begin{center}
\begin{tabular}{cc}
\includegraphics*[bb=16 18 278 220, width=0.46\textwidth]{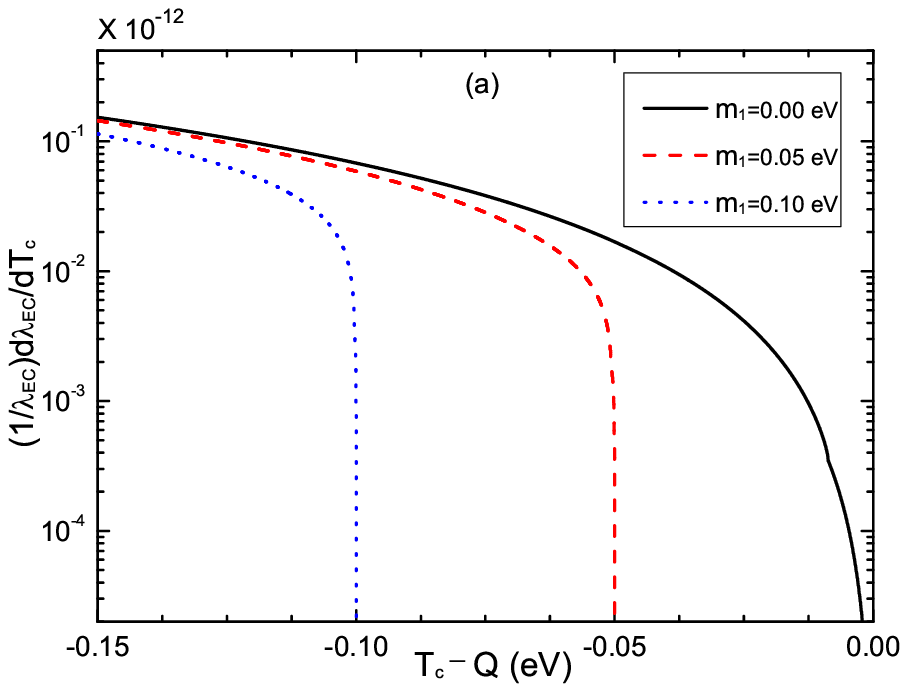}
&
\includegraphics*[bb=16 18 278 220, width=0.46\textwidth]{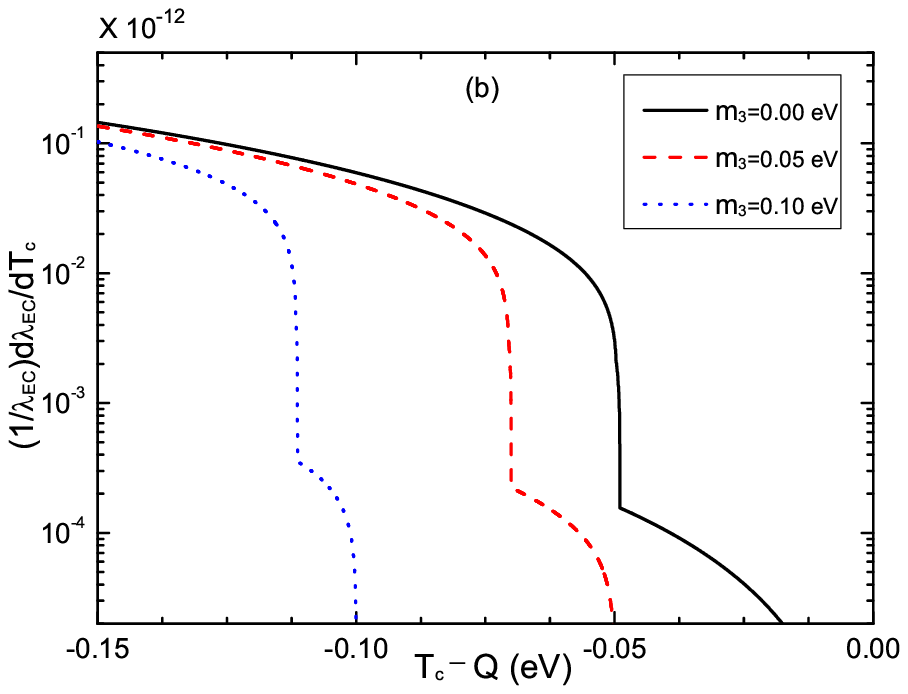}
\\
\end{tabular}
\end{center}
\vspace{-0.5cm}
\caption{The fine structure spectrum near the endpoint of the
$^{163}{\rm Ho}$ EC decay in the $m^{2}_{31}>0$ (left panel) or
$m^{2}_{31}<0$ (right panel) case.}
\end{figure}
\begin{figure}[t]
\begin{center}
\begin{tabular}{cc}
\includegraphics*[bb=18 16 276 210, width=0.46\textwidth]{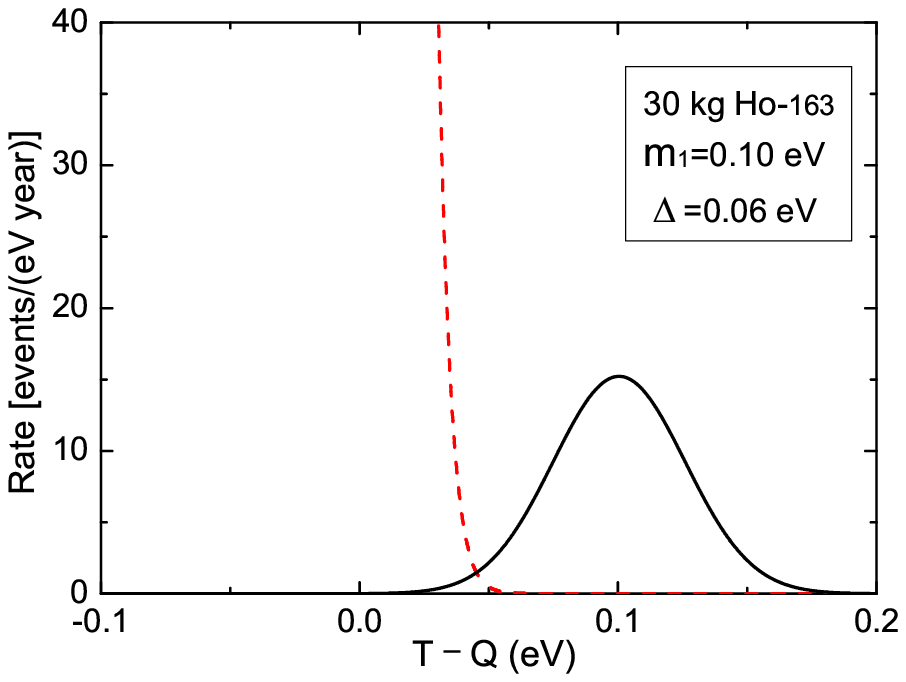}
&
\includegraphics*[bb=18 18 277 214, width=0.46\textwidth]{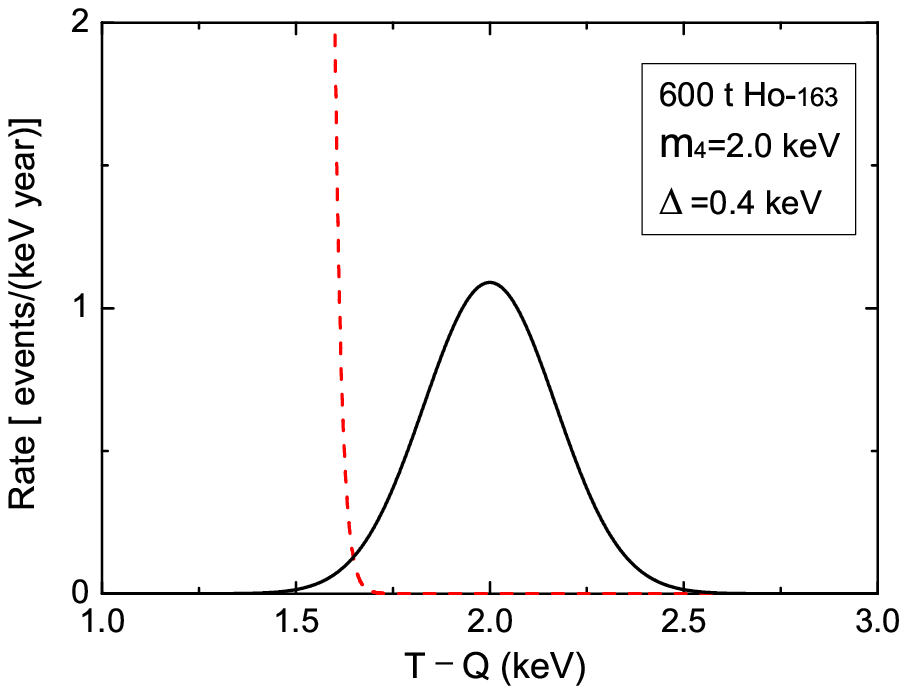}
\end{tabular}
\end{center}
\vspace{-0.5cm}
\caption{The antineutrino capture rate as a function of the overall
energy release for hot DM antineutrinos (left panel) and warm DM
antineutrinos (right panel).}
\end{figure}

\section{Concluding Remarks}

To pin down what DM is really made of has been one of the most important
and most challenging problems in particle physics and cosmology.
Both the active and sterile species of sub-eV
neutrinos and antineutrinos might be a part of hot DM,
and keV sterile neutrinos and antineutrinos could be a good
candidate for warm DM. Here we have addressed ourselves to
the direct laboratory detection of possible contributions of relic
neutrinos and antineutrinos to DM. The beta-decaying (e.g.,$^3{\rm H}$
and $^{106}{\rm Ru}$) and EC-decaying (e.g., $^{163}{\rm Ho}$)
nuclei have been considered as the most promising targets to capture such
low-energy neutrinos and antineutrinos, respectively.
Our analysis shows that the
signatures of hot or warm DM neutrinos and antineutrinos should in
principle be observable, provided the target is big enough, the
energy resolution is good enough and the gravitational clustering
effect is significant enough. We admit that our numerical results
are quite preliminary and mainly serve for illustration, but we
stress that such a direct laboratory search for hot or warm DM
neutrinos and antineutrinos is fundamentally important and deserves
further attention and more detailed investigations. Although the
present experimental techniques are unable to lead us to a
guaranteed measurement of relic neutrinos and antineutrinos in the
near future, we might have a chance to make a success of this great
exploration in the long term.

\section{Acknowledgements}

One of us (Z.Z.X.) would like to thank M. Biesiada for his
kind invitation and warm hospitality in Ustron, where this wonderful
conference was held.
This work was supported in part by the China Postdoctoral Science
Foundation under grant No. 20100480025 (Y.F.L.) and in part by the
National Natural Science Foundation of China under grant No. 10875131
(Z.Z.X.).

\end{document}